\documentclass[aps,twocolumn,prl,groupedaddress,showkeys]{revtex4-1}
\usepackage{graphicx}
\usepackage{booktabs}
\usepackage[utf8]{inputenc}
\usepackage{wrapfig}
\usepackage{mathtools}
\usepackage[colorlinks=true,linkcolor=blue]{hyperref}

\begin{document}

\title{Photo-induced reversible modification of the Curie-Weiss temperature in paramagnetic gadolinium compounds} 

\author{{José~Montero\textsuperscript{$\star$}, Peter Svedlindh, Lars Österlund\\}{Department of Materials Science and Engineering \unskip, Uppsala University\unskip, Uppsala\unskip, (Sweden).}\\ 
\textsuperscript{$\star$}email : jose.montero-amenedo@angstrom.uu.se. }
\date{\today}

\begin{abstract}
Gadolinium oxyhydride GdHO is a photochromic material that darkens under illumination and bleaches back by thermal relaxation. As an inorganic photochromic material that can be easily deposited by magnetron sputtering, GdHO has  very interesting potential applications as a functional material, specially for smart glazing applications.  However, the underlying reasons behind the photochromic mechanism--which can be instrumental for the correct optimisation of GdOH for  different applications--are not completely understood. In this paper, we rely on the well-stablished magnetic properties of Gd$^{3+}$ to shed  light on this matter. GdOH thin films present paramagnetic behaviour similar to other Gd$^{3+}$ compounds such as Gd$_2$O$_3$. Illumination of the films result in a reversible increase of the Curie-Weiss temperature pointing to RKKY interactions,  which is consistent with the resistivity decrease observed in the photo-darkened films.

\end{abstract}

\flushbottom
\maketitle

\thispagestyle{empty}

Single anion compounds, e.g. oxides, nitrides, fluorides, have been extensively studied during the past decades and play a cornerstone role in today's technological development. In contrast, multiple-anion compounds, that is, materials in which two or more anions coexist in a single phase, e.g. oxyhydrides, have been barely explored \cite{Kageyama2018}. 
Some notable examples of oxyhydrides include LaHO \cite{Ooya2021, Malaman1984}, NdHO \cite{Kutsuzawa2019,Wideroee2011}, YHO \cite{Zapp2019} and GdHO \cite{Ueda2018}.
According to Kageyama \emph{et al.} \cite{Kageyama2018}, due to the flexibility resulting from the introduction of different anions, oxyhydrides (and mixed anions in general)  can constitute a scaffolding upon which future functional materials exhibiting \emph{new or enhanced properties} [sic] can be built.
This idea is supported by  the rare-earth (e.g. Dy, Er, Gd, La, Y) oxyhydride family , which has arisen in recent years as a new group of materials exhibiting photochromic \cite{Nafezarefi2017}, luminescent \cite{Ueda2018} and photocatalytic properties \cite{Ooya2021}. 

Rare-earth oxyhydrides are  positive  T-type  photochromic materials that is, they darken upon illumination and bleach by thermal relaxation. Photochromism in oxyhydrides has  attracted attention due to its potential industrial applications for, e.g., the fabrication of photochromic smart windows, given their superiority in certain technological aspects when compared with conventional photochromic materials.
Photochromic oxyhydrides can be easily fabricated in thin film form by magnetron sputtering \cite{Montero2018}, which gives them a huge advantage over the popular photochromic silver halides and, in addition, are inorganic and thus presumably more resilient to harsh conditions (particularly to high UV doses) than the photochromic organic dyes commonly used by the industry nowadays \cite{Towns2016}.

Understanding the cause of the photochromic response in rare-earth oxyhydrides is of crucial importance towards a practical application of such compounds. Unfortunately, the underlying physico-chemical reasons behind the photochromic mechanism are not completely understood. There is, however, a consensus that points to the formation during illumination of metallic-like domains in the oxyhydride lattice \cite{Montero2017,Komatsu2022, Wu2022}, a phenomenon that is accompanied by persistent photoconductivity \cite{You2019}. It is also known that the lattice of these oxyhydrides contracts under illumination, and it relaxes back to the original state during bleaching \cite{Baba2020}. 

At this point, it is natural to wonder how these structural and electronic changes induced by illumination affect the magnetic properties of rare-earth oxyhydride thin films  and what can we learn about the photochromic mechanism from a magnetic perspective.
Accordingly, among different rare-earths oxyhydrides, in this work we have selected GdOH. Our choice is based on the electronic configuration of Gd$^{3+}$, i.e., $[$Xe$]4f^7$, which endows Gd$^{3+}$ with a large magnetic moment. If we assume Gd$^{3+}$ to govern the magnetic behaviour of  gadolinium oxyhydride compounds -- of stoichiometry GdO$_x$H$_y$ but to which we will refer in the text as GdOH for simplicity -- then we expect a paramagnetic behaviour and a reasonable agreement to the Curie-Weiss law. This paper  aims to answer the following questions: \emph{(i)} how does the magnetic susceptibility $\chi$ of GdOH thin films looks like and how does it compare with $\chi$ corresponding to other closely related compounds, namely GdH$_2$  and Gd$_2$O$_3$? \emph{(ii)} How does illumination affects $\chi$ in GdOH and what can we learn about the photochromic mechanism from it?

For our purpose, GdHO thin films were prepared by a two-stage synthesis process as described elsewhere \cite{Nafezarefi2017, Montero2018}. This process comprises the deposition of GdH$_2$ films by reactive magnetron sputtering onto glass substrates under conditions that result in a porous structure that allows for the quick incorporation of oxygen, and hence the formation of GdOH, once the films are exposed to air \cite{Zubkins2022}.  In our case the conditions were 8.6$\times10^{-7}$ mbar for base pressure, 6$\times10^{-3}$ mbar for working pressure (40 sccm and 12 sccm Ar and H$_2$ flow, respectively) and 100 W power applied to a 2 inch. diameter Gd target (TREM 99.6 \%, Gd/TREM 99.9 \%).  The thickness of each sample was determined by a Bruker DektakXT stylus profilometer using the step between film and substrate.

The magnetic susceptibility of the obtained GdOH films (before and after illumination) was determined in a Magnetic Properties Measurement System MPMS (Quantum Design), between $\sim$ 5 K and $\sim$ 150 K in a field of 50 mT. For this measurement the films were cut in pieces of area $\sim$ 0.4$\times$0.4 cm$^2$. The uncoated substrate (glass) was measured in the same temperature range applying  1 T and  the obtained results used for the diamagnetic correction of the measurements.  For  the optical and structural properties of the GdOH films before and after illumination see Supplemental Material \cite{supplemental}, which includes Refs \cite{Colombi2020,Baba2020, Li2013}. 

\begin{figure} 
\centering
\includegraphics[width=0.5\textwidth]{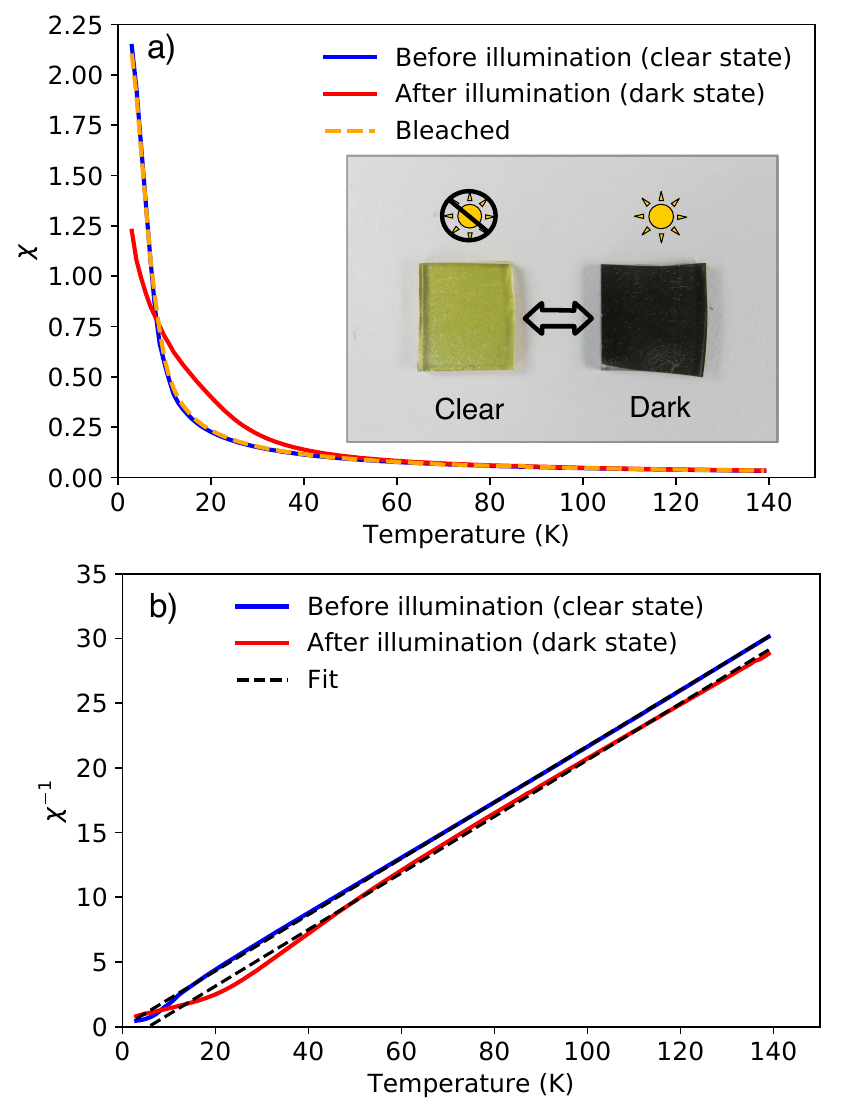}
\caption{Panel (a): susceptibility $\chi$ as a function of temperature $T$ for a GdOH sample before (clear) and after (dark) 15 h illumination with a 365 nm-light source (27 mW/cm$^2$). The change is reversible as shown by the curve corresponding to the bleached state. Inset in (a) shows the photograph of two fragments of the same GdOH sample before (left) and after (right) illumination. Panel (b): inverse of the magnetic susceptibility $\chi^{-1}$ vs. $T$ in the clear and dark states together with their respective fits according to the Curie-Weiss law.}
\label{fig1}
\end{figure}

Figure \ref{fig1} (a) depicts the magnetic susceptibility $\chi$ corresponding to a clear, photodarkened (dark state) and recovered (bleached) GdHO sample (approx. 1.0 $\mu \mathrm{m} $-thick). The photodarkened state was obtained after 15 h illumination (wavelength $\lambda = 365\; \mathrm{nm}$, power 27 mW/cm$^2$) and the bleached state was measured after allowing the films to bleach during several weeks, knowing that the actual bleaching time is considerably shorter \cite{Colombi2021}.  Inset in Figure \ref{fig1} (a) shows photographs of two fragments of the same GdHO sample before and after illumination. The change in luminous transmittance of the film before/after illumination  is $\sim$ 50 \% \cite{supplemental}.
Since susceptibility studies of the photodarkened state have to be done \emph{ex-situ}, risk of partial bleaching during measurement exists. However, we assume that in a helium gas atmosphere and at low temperature (the experimental conditions) the recovery is negligible. In any case, $\chi$ vs. $T$ curves  were measured from low to high $T$ and from high to low $T$, resulting in  perfect match, which confirms that the photodarkened sample has not changed during the measurement. The data presented in Figure \ref{fig1}  corresponds to measurements performed \emph{exactly} in the same fragment of sample in the clear, dark and bleached state.

From Fig. \ref{fig1} (a) we can infer that \emph{(i)} gadolinium oxyhydride shows a paramagnetic behaviour and \emph{(ii)}, illumination affects $\chi$ in gadolinium oxyhydride, specially at low temperatures. It is also important to note that the change is reversible, since the $\chi$ vs. $T$ curve recovers its initial shape after bleaching. 

According to the Curie-Weiss law, $\chi$ can be written as a function of the temperature $T$ as follows:
\begin{equation}
\chi = \frac{C}{T-\Theta_{CW}},
\label{eq1}
\end{equation}
being $C$ the Curie constant and $\Theta_{CW}$ the Curie-Weiss temperature.  $C$ is related to the effective magnetic moment per ion $\mu_{\mathrm{eff}}$ by:
\begin{equation}
\mu_{\mathrm{eff}}= \sqrt{\frac{3k_B}{\mu_0N}}\cdot\frac{1}{\mu_B}\cdot\sqrt{C},
\label{eq2}
\end{equation}
where $k_B$ is the Boltzmann constant, $\mu_0$ the vacuum magnetic permeability, $\mu_B$  the Bohr magneton and $N$ the number or magnetic ions (i.e., in our case Gd$^{3+}$) contained in the sample.

The parameters $C$ and $\Theta_{CW}$ can, according to Equation \ref{eq1}, be obtained by fitting $\chi^{-1}$ vs. $T$, see Figure \ref{fig1} (b).  The curve corresponding to the bleached sample is not shown in Figure \ref{fig1} (b) for simplification, since it overlaps with the one corresponding to the clear state. Note that we have selected to fit $\chi^{-1}$ vs. $T$ rather than $\chi$ vs. $T$. In this way the weight of the fitting moves towards higher temperatures where the compliance with the Curie-Weiss law is best \cite{Mugiraneza2022}.  In fact, we can observe in Figure \ref{fig1} (b) how $\chi$ deviates from the Curie-Weiss law at low temperatures, the deviation being more important when the sample is in the photodarkened state. 

 The reversible increase observed in $\Theta_{CW}$ in the photodarkened samples (ferromagnetic interaction) can be tentatively explained by the increase in electrical conductivity after illumination reported in GdHO films \cite{Colombi2023}, and hence the possible exchange interaction between conduction  electrons and \emph{4f} electrons according to the RKKY interaction model, as often reported in GdH$_x$ compounds \cite{Yayama1992}. This is consistent with the study of the sheet resistance of our samples as performed in an AIT-CMT-SR200N four-point prove instrument before and after illumination. Before illumination, the sheet resistance corresponding 1.0 $\mu$m-thick GdOH films was out of range of our instrument (which according to the manufacturer's specifications, corresponds to values above 2 MOhm/sq). In the photodarkened films, however, the sheet resistance dropped to the kOhm/sq range, although repeated measurements presented a large standard deviation (7.0 $\pm$ 3.0 kOhm/sq).

$N$ can be estimated from the volume of each sample and the density of GdOH. Thus, by inserting $N$ in Equation \ref{eq2}, we can calculate $\mu_{\mathrm{eff}}$ from the value of $C$ obtained from the fitting shown in Figure \ref{fig1} (b). However, the situation is not  straightforward because the samples are thin films prepared by sputtering in a porous deposition regime, and hence less dense than bulk. Moreover GdOH, due to its mixed-anion nature, has a flexible stoichiometry and, what is worse, can contain other phases (particularly amorphous Gd$_2$O$_3$). In any case, we can assume that the actual density of our samples must lie between that of GdH$_2$ and Gd$_2$O$_3$.  
Accordingly, in Table \ref{t1} we report $\mu_{\mathrm{eff}}$ calculated using the density of Gd$_2$O$_3$  and GdH$_2$ taken  from the data contained in their respective International Centre of Diffraction Data ICDD-pdf cards num. 00-120797 and 00-050-1107 and   the density estimated for GdHO by Colombi \emph{et al.} \cite{Colombi2020}. 
Irrespectively of the density chosen, the obtained values of $\mu_{\mathrm{eff}}$ are in reasonable agreement with experimental ($\mu_{\mathrm{eff}}$ = 7.90 $\mu_B$) and theoretical  ($\mu_{\mathrm{eff}}$ = 7.94 $\mu_B$) values for Gd$^{3+}$ \cite{Mugiraneza2022}.
 
 The data presented in Figure \ref{fig1} corresponds to a sample of  approximate thickness $d$ of 1.0 $\mu m$,  labeled as S1 in Table \ref{t1}.  The data corresponding to an additional sample, S2, of thickness $d\sim0.5$ $ \mu m$ is also presented in Table \ref{t1}. The data corresponding to the photodarkened state was collected after irradiating S1 and S2 with a 365 nm-light source (27 mW/cm$^2$) during 15 and 19 h, respectively.  Note that, contrary to the case of S1, the clear and dark state of S2 have not been measured in the same fragment, but in different fragments of the same sample, and hence errors arising from different areas and sample inhomogeneity may occur.   
 
 Over 16 measurements were performed on  different samples under different illumination conditions, always with the same outcome: increase of $\Theta_{CW}$ after illumination.  A preliminary observation suggests that there is a direct relation between illumination time and change in  $\Theta_{CW}$. However the experimental limitations of  the present study-- inhomogeneity of lamp beam and  sample size differences-- prevent us from making any definite conclusion in this respect. All measurements confirm qualitatively the results presented in Figure \ref{fig1} and Table \ref{t1}, namely the increase in $\Theta_{CW}$ after illumination.


\begin{table} 

\centering
\begin{center} 
\begin{tabular}{|c|p{1.2cm}|p{1.2cm}|p{1.2cm}|p{1.2cm}|p{1.2cm}|}

\toprule
\hline

Sample & $d$ ($\mu$m) &$\Theta_{CW}$ (K) & $\mu_{\mathrm{eff}}^\ast$   & $\mu_{\mathrm{eff}}^\dagger$  & $\mu_{\mathrm{eff}}^\ddagger$  \\
\hline
\hline
S1-Clear & 1.0 &0.15& 8.46& 8.13 &  8.35  \\ 
\hline
S1-Dark & 1.0 &5.63& 8.44& 8.10 &  8.33  \\ 
\hline
S1-Bleach & 1.0 &0.19& 8.53& 8.19 &  8.42  \\ 
\hline
\hline
S2-Clear & 0.5 &0.37& 8.1& 7.78 &  8.00  \\ 
\hline
S2-Dark & 0.5 &10.36& 7.99& 7.67 &  7.89  \\ 
\hline
\end{tabular}
\caption{ Thickness $d$, Curie-Weiss temperature $\Theta_{CW}$ and effective magnetic moment per ion $\mu_{\mathrm{eff}}$ for two gadolinium oxyhydride samples (S1 and S2)  in the clear, dark and bleached state. $\mu_{\mathrm{eff}}$ is given in units of Bohr magnetons, and has been calculated assuming the density of the sample equal to the one of Gd$_2$O$_3$ ($^\ast$), GdH$_2$ ($^\dagger$)  and GdHO ($^\ddagger$). }
\label{t1}
\end{center}
\end{table}

\begin{figure} 
\centering
\includegraphics[width=0.5\textwidth]{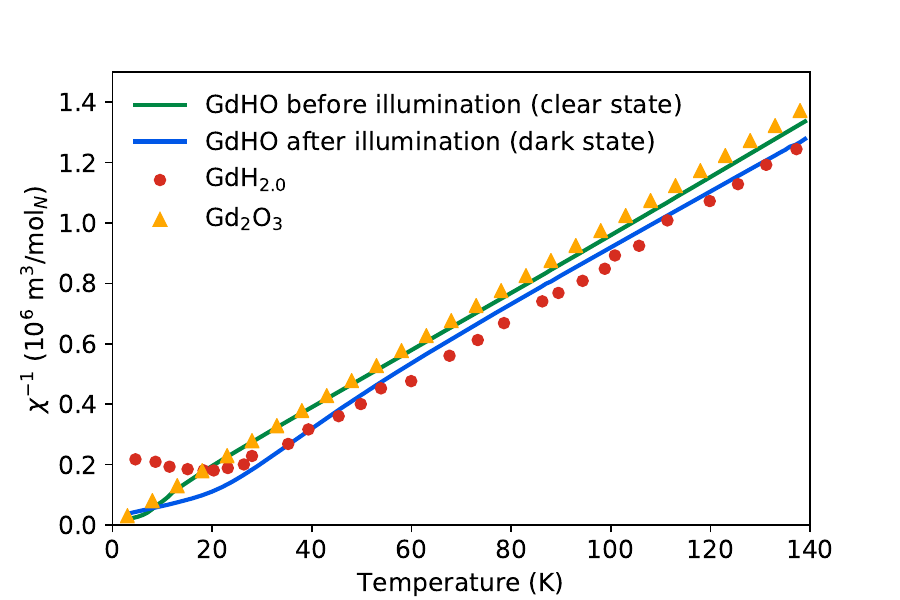}
\caption{ $\chi^{-1}$ corresponding a GHO thin film in the clear and dark state, compared to data corresponding to Gd$_2$O$_3$ and GH$_{2.01}$ reported by Mugiraneza  \emph{et al.} \cite{Mugiraneza2022} and Wallace \emph{et al.} \cite{Wallace1963}, respectively.}
\label{fig2}
\end{figure}

In Figure \ref{fig2}, $\chi^{-1}$ vs $T$  curves corresponding to GdHO (sample S2) in the clear and photodarkened state are compared to those of Gd$_2$O$_3$ and GH$_{2.01}$ as reported by Mugiraneza  \emph{et al.} \cite{Mugiraneza2022} and Wallace \emph{et al.} 
\cite{Wallace1963}, respectively. Note that for convenience when comparing with literature data, the susceptibility in Figure \ref{fig2} is given in molar units, being mol$_N$ the product of the number of moles  times the number of magnetic ions per formula unit \cite{Mugiraneza2022}.

Figure \ref{fig2} is conveniently discussed  in light of the magnetic properties of the Gd--H system \cite{Wallace1963}: metallic Gd exhibits strong ferromagnetic properties, whereas fully hydrogenated GdH$_3$ exhibits no tendency to magnetic ordering. Thus, according to Wallace \emph{et al.} \cite{Wallace1963}, the interactions between \emph{4f} electrons must occur trough delocalised conduction electrons which, while abundant in metallic Gd, are absent  in fully hydrogenated GdH$_3$. Similarly to GdH$_3$, the oxide Gd$_2$O$_3$ has no conduction electrons \cite{Mugiraneza2022} and thus behaves as a paramagnet, see Figure \ref{fig2}.
As ferromagnetic Gd metal is hydrogenated and Gd hydride is formed, conducting electrons are trapped in the hydride anion H$^-$. Consequently, the free electron concentration is reduced, which causes a decrease of  the saturated magnetic moment \cite{Yayama1992}. 
At stoichiometries GdH$_x$, with $x$ in the range $1.8\leq x \leq 2.3$, although from an optical perspective the material still behaves as a metal \cite{Colombi2020}, the behaviour is very different from a magnetic point of view. In this stoichiometry range an antiferromagnetic transition is usually reported at a temperature $T_N \sim 20$ K \cite{Wallace1963, Yayama1992}, above which the the material follows the Curie-Weiss law, as illustrated in Figure \ref{fig2} by GdH$_2$.  In particular, for the case of GdH$_{2.0}$, Wallace \emph{et al.} \cite{Wallace1963} report a Neél temperature $T_N$ = 21 K, $\Theta_{CW}$ = 3.0 K and $\mu_{\mathrm{eff}}$ = 7.7 $\mu_B$ (our analysis carried out in their digitised data resulted in $\mu_{\mathrm{eff}}$ = 7.9 $\mu_B$ and $\Theta_{CW}$ = 6.0 K). Our study of the digitised data provided by Yayama \emph{et al.} \cite{Yayama1992} for GdH$_2$ (not shown), resulted in  $\mu_{\mathrm{eff}}$ = 7.9 $\mu_B$ and $\Theta_{CW}$ = 6.20 K.
We note that, although Gd cations in GH$_2$  are in the oxidation state Gd$^{2+}$,  $\mu_{\mathrm{eff}}$ is  governed by the $4f$ electrons, whereas the extra delocalised electron is the cause of exchange  interactions that affect $\Theta_{CW}$.


In light of the study of the literature, our results can be interpreted as follows: before illumination GdHO is highly resistive and contains no free electrons \cite{Colombi2023, Miniotas2000}, thus  behaving as a paramagnet similar to Gd$_2$O$_3$, Figure \ref{fig2}.  
On the other hand, as observed by Colombi \cite{Colombi2023}, and in agreement with our sheet resistance measurements, the free electron concentration is larger in photodarkened GdHO films, and  hence the emergence of ferromagnetic interactions reflected in the increase of $\Theta_{CW}$ presented in Table \ref{t1}.
 As a consequence, the curve $\chi^{-1}$ vs. $T$  corresponding to GdHO in the photodarkened state is similar to the one corresponding to GdH$_2$, Figure \ref{fig2}. As stated before, this is because in both GH$_2$ and photodarkened GdHO the magnetic properties are governed by the $4f$ electrons, although the exchange interactions caused by the conducting electrons need to be considered. We note that the results from temperature dependent susceptibility measurements for the photodarkened state do not reveal the signature of an antiferromagnetic transition, instead a broad feature in $\chi$ vs. $T$ is observed (cf. Figure \ref{fig1} (a)). A likely explanation for this is that the RKKY interactions, considering both nearest neighbour and next-nearest neighbour interactions, include both ferro- and antiferromagnetic interactions. The existence of both ferro- and antiferromagnetic interactions may explain why previous studies of GdH$_{2.0}$ report an antiferromagnetic transition at the same time as reporting a positive $\Theta_{CW}$ value \cite{Wallace1963}.

 A likely cause of the increase of free electron density after illumination in GdHO is the reduction of some Gd$^{3+}$ cations to Gd$^{2+}$ \cite{Colombi2023}. 
 This hypothesis is in agreement with our observations, since it will allow  the slope of the $\chi^{-1}$ vs. T curve to remain unaltered before and after illumination.  The photoreduction of Gd$^{3+}$ to Gd$^{2+}$ would result in an increase of $\Theta_{CW}$ in the dark state, which explains the observations in Figures \ref{fig1}, \ref{fig2}, and Table \ref{t1}. We cannot rule out, however, other causes behind the increment of free carriers after illumination.

\section*{Conclusions}

The magnetic susceptibility of the GdHO films in the clear state correspond to that of a paramagnetic material, very similar to Gd$_2$O$_3$, with an effective magnetic moment per ion of $\sim$ 7.9 $\mu_B$, as expected for Gd$^{3+}$ compounds. 
Illumination causes the emergence of persistent magnetic exchange interactions,  which lead to  an increase of the Curie-Weiss temperature when compared to the initial transparent state.
These magnetic exchange interactions, which disappear when the GdHO films are allowed to relax back to the transparent state,
point to the formation of metallic domains occurring upon photorreduction of some Gd$^{3+}$ cations into Gd$^{2+}$. However, the formation of Gd$^{2+}$ cannot be directly inferred from the determination of the effective magnetic moment which is governed by the $4f$ electrons in both Gd$^{3+}$ and Gd$^{2+}$.

\section*{Acknowledgements}

This work has been supported by the Swedish Energy Agency E2B2 Project Ref. P2022-00859. J. M. is grateful to Dr. H. Stopfel for helping in operating the MPMS and to Dr. R. Mathieu for the fruitful discussions. 

\section*{}
\bibliography{References_Magnetic_Manuscript.bib}
\bibliographystyle{apsrev4-1}


\end{document}


\title{Supplemental Material: Photo-induced reversible modification of the Curie-Weiss temperature in paramagnetic gadolinium compounds} 
\author{{José~Montero\textsuperscript{$\star$}, Peter Svedlindh, Lars Österlund\\}{Department of Materials Science and Engineering \unskip, Uppsala University\unskip, Uppsala\unskip, (Sweden).}\\ 
\textsuperscript{$\star$}email : jose.montero-amenedo@angstrom.uu.se. }
\date{\today}

\maketitle
The crystalline structure of photochromic GdOH films was studied by grazing incidence x-ray diffraction (GIXRD) in a Siemens D5000 diffractometer using Cu K$\alpha$ radiation ($\lambda$ = 1.5418 Å). The angle of incidence was fixed to 1$^\circ$. 

\begin{figure}
\centering
\includegraphics[width=0.45\textwidth]{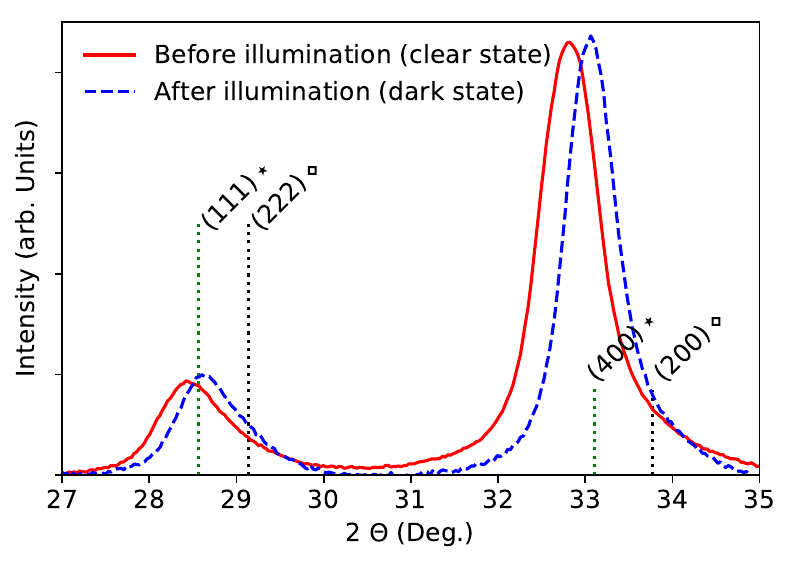}
\caption{X-ray diffraction patterns corresponding to a GdHO sample (1 $\mu$m-thick, labeled S1 in the main text) before and after illumination: 15 h illumination (wavelength $\lambda = 365\; \mathrm{nm}$, power 27 mW/cm$^2$). Vertical lines correspond to diffraction peaks according to  ICDD PDF 00-050-1107 (for GdH$_{1.8}$, labeled  $\diamond$ ) and ICDD-PDF 00-12-0797 ( for Gd$_2$O$_3$, labeled $\star$) with Miller indices in parenthesis. }
\label{fig_xrd} 
\end{figure}

Diffractograms for the sample in the clear and dark state are shown in Figure \ref{fig_xrd}.  Vertical lines in Figure \ref{fig_xrd} represent standard position of the diffraction peaks as reported by the International Centre for Diffraction Data (ICDD) PDF 00-050-1107 corresponding to GdH$_{1.8}$, as well as  ICDD-PDF 00-12-0797 corresponding to Gd$_2$O$_3$ (labeled $\diamond$ and $\star$ in the figure, respectively). Miller indices are indicated in parenthesis. The structure of GdHO can be matched with the \emph{fcc} symmetry $Fm\overline{3}m$ with diffraction peaks indicating larger interplanar distances that those in Gd$_2$O$_3$ or in GdH$_{1.8}$ \cite{Colombi2020}. 

In Figure \ref{fig_xrd} we can observe how illumination causes the reversible contraction of the lattice, resulting in a  displacement of the diffraction peaks towards larger angles. We have observed the same effect in photochromic YHO \cite{Baba2020}.  
 
Additionally, the optical transmittance T of the GdHO films was measured in the UV-VIS-NIR range before and after illumination using a spectrophotometer Perkin-Elmer Lambda 900 equipped with an integrating sphere. 

\begin{figure}[h] 
\centering
\includegraphics[width=0.45\textwidth]{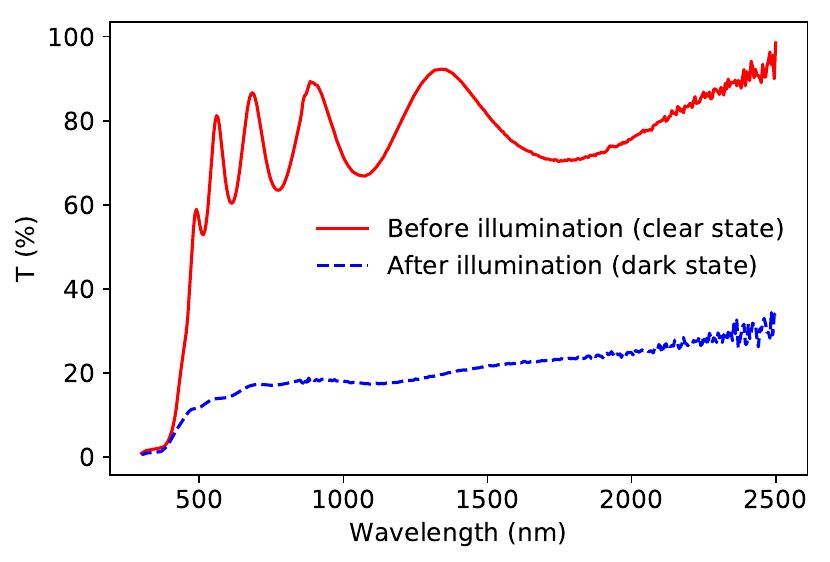}
\caption{Transmittance T corresponding to a GdOH film (1 $\mu$m-thick, labeled S1 in the main text) before and after 15 h illumination (wavelength $\lambda = 365\; \mathrm{nm}$, power 27 mW/cm$^2$).  }
\label{fig_Optics} 
\end{figure}

Figure \ref{fig_Optics}  depicts T curves of a GdOH sample. After illumination, the transmittance of the sample decreased notably. In fact, the solar transmittance, calculated as described in Ref. \cite{Li2013} is 66  \% and 13  \% for the sample in the clear and dark state, respectively, which results in a modulation of the solar irradiance above 50 \%. Analogously, the the luminous transmittance, calculated as described in Ref. \cite{Li2013}  is 65  \% in the clear state and 16 \% in the dark state.

\section*{}
\bibliography{References_Magnetic_Manuscript.bib}
\bibliographystyle{apsrev4-1}